\documentclass[twocolumn]{emulateapj}
\usepackage{graphicx}
\usepackage{dcolumn}

\graphicspath{{./Users/nathansecrest/Documents/Papers/ngc4178agn/}}

\def\arcsec{\hbox{$^{\prime\prime}$}}
\def\deg{\hbox{$^\circ$}}
\def\init{\hspace{0.75 mm}}

\defcitealias{secrest12}{S12}

\begin{document}

\title{A Multi-Wavelength Analysis of NGC 4178: A Bulgeless Galaxy with an AGN}

\author{N.~J.~Secrest\altaffilmark{1}, S.~Satyapal\altaffilmark{1}, S.~M.~Moran\altaffilmark{2}, C.~C.~Cheung\altaffilmark{3}, M.~Giroletti\altaffilmark{4}, M.~Gliozzi\altaffilmark{1}, M.~P.~Bergmann\altaffilmark{5} \& A.~C.~Seth\altaffilmark{6}}

\altaffiltext{1}{George Mason University, Department of Physics \& Astronomy, MS 3F3, 4400 University Drive, Fairfax, VA 22030, USA}

\altaffiltext{2}{Johns Hopkins University, Department of Physics \& Astronomy, 366 Bloomberg Center, 3400 N. Charles Street, Baltimore, MD 21218, USA}

\altaffiltext{3}{Space Science Division, Naval Research Laboratory, Washington,
DC 20375-5352, USA}

\altaffiltext{4}{INAF Istituto di Radioastronomia, 40129 Bologna, Italy}

\altaffiltext{5}{NOAO Gemini Science Center (Chile), P.O. Box 26732, Tucson, AZ 85726}

\altaffiltext{6}{Department of Physics and Astronomy, University of Utah, Salt Lake City, UT 84112, USA}

\begin{abstract}
We present {\it Gemini}  longslit optical spectroscopy and VLA radio observations of the nuclear region of NGC~4178, a late-type bulgeless disk galaxy recently confirmed to host an AGN through infrared and X-ray observations.  Our observations reveal that the dynamical center of the galaxy is coincident with the location of the {\it Chandra} X-ray point source discovered in a previous work, providing further support for the presence of an AGN.  While the X-ray and IR observations provide robust evidence for an AGN, the optical spectrum shows no evidence for the AGN, underscoring the need for the penetrative power of mid-IR and X-ray observations in finding buried or weak AGNs in this class of galaxy.  Finally, the upper limit to the radio flux, together with our previous X-ray and IR results, is consistent with the scenario in which NGC~4178 harbors a deeply buried AGN accreting at a high rate.

\end{abstract}

\section{Introduction}

The fact that supermassive black holes (SMBHs, $M_{\mathrm{BH}}\gtrsim10^{6}$~M$_{\sun}$) or intermediate-mass black holes (IMBHs, $10^{2}$~M$_\sun$~$\lesssim M_{\mathrm{BH}}\lesssim10^{6}$~M$_{\sun}$) can be found in late-type galaxies in which there is no evidence for classical bulge is now on firm empirical ground~\citep{filippenko03,barth04,greene04,greene07,satyapal07,dewangen08,ghosh08,mathur08,satyapal08,shields08,barth09,desroches09,gliozzi09,satyapal09,jiang11a,jiang11b,mcalpine11,secrest12,arayasalvo12,simmons13}.  The presence of such black holes in these galaxies points to a new understanding of the formation mechanisms of large nuclear black holes. This is further bolstered by the remarkable evidence for the presence of an active galactic nucleus (AGN) in the dwarf galaxy Henize~2-10~\citep{reines11} and the mounting evidence for AGN activity in blue compact dwarf galaxies~\citep{izotov07,izotov08,izotov10}.  It is also likely that the particularly bright ultraluminous X-ray source HLX-1, also an IMBH~\citep{soria10,webb12}, is the remnant of a dwarf galaxy~\citep{farrell12}.  Central to this new understanding is the black hole occupation fraction, which can provide insights into differentiating whether the first `seed' black holes in the Universe originated from Population III stars, or whether the first seed black holes had a more exotic origin, such as the hypothetical `quasistar' direct-collapse scenario~\citep[see][and references therein]{bonoli12} that may have unfolded in the early Universe due to the extremely low metallicity environment and lack of stellar clumping~\citep[for details on how the black hole occupation fraction differs under these two scenarios, see, e.g.,][]{vanWassenhove10}.  These bulgeless and dwarf galaxies are of particular importance to this question because classical bulges are thought to form through mergers, which erase information on the progenitor galaxies' black hole masses after they coalesce, while pseudobulges are thought to form through secular processes~\citep{kormendy04}.  Nuclear black holes in these galaxies may thus be thought of as `pristine', having likely gone through only secular growth during the extent of cosmic history, and preserving the characteristics of the seed population from which they originated.  While it is not yet technically feasible to obtain an unbiased estimate of the black hole occupation fraction at masses $\lesssim10^{6}$~M$_{\sun}$ due to limitations on detectability\footnote{Although see~\citet{greene12} for a first-pass estimate, which tentatively finds that the direct-collapse scenario may better explain the (limited) data we currently have on this black hole demographic.}, we can nonetheless begin to build a census of AGNs in low mass and bulgeless galaxies that will contribute to our understanding of this subject.

One such galaxy is NGC~4178, a late-type, bulgeless spiral galaxy ($d=16.8$~Mpc) that was recently confirmed to host an AGN through high-resolution {\it Chandra} X-ray observations~\citep[hereafter S12]{secrest12}, with a nuclear black hole mass, inferred through mid-IR and radio/X-ray correlations, between $\sim10^{4}-10^{5}$~M$_{\sun}$.  While the presence of an AGN has been confirmed in this object, the optical confirmation of the presence of an AGN in NGC 3621~\citep{barth09}, a bulgeless spiral galaxy with highly similar X-ray/mid-IR properties, has motivated us to explore the optical properties of NGC~4178.  While an optical spectrum of the nuclear region already exists~\citep{ho95}, we have determined that it did not actually cover the region containing the nuclear X-ray source, and so it is not known how the ionization of the interstellar medium (ISM) changes near the source, if at all.  Furthermore, it is not definitively known whether or not some of the [Ne~V] emission from the nucleus~\citep{satyapal09} originates in a starburst and/or shocks.  If this were true, then it would reduce the inferred bolometric luminosity of the AGN and therefore the lower limit on the mass of the nuclear black hole.  By examining the star formation properties of the nuclear region, we can explore this possibility and make the inferred bolometric luminosity more robust.

The aim of this paper is to further characterize the nuclear environment around the AGN in NGC~4178.  We used {\it Gemini} longslit spectroscopy to derive the dynamical center of the galaxy, explore the ionization state of the circumnuclear ISM, and explore the presence of star formation.  We also obtained new Very Large Array (VLA) observations to further constrain the mass of the nuclear black hole.  In \S{2}, we describe the {\it Gemini} observations and data reduction, as well as the new VLA data.  We continue with a description of our results in \S{3}, before discussing the implications of our results in \S{4}.  We give a summary and our primary conclusions in \S{5}.

\section{Observations and Data Reduction}

\subsection{{\it Gemini} Spectroscopic Observations}

NGC 4178 was observed with the {\it Gemini} GMOS-N, long-slit ($1\arcsec$), B600-G5307 grating on 2012 February 14th and 21st for 3600 seconds and 2400 seconds (1200 seconds per frame), respectively (Program ID: GN-2012A-Q-14)\footnote{Based on observations obtained at the Gemini Observatory, which is operated by the Association of Universities for Research in Astronomy, Inc., under a cooperative agreement with the NSF on behalf of the Gemini partnership: the National Science Foundation (United States), the Science and Technology Facilities Council (United Kingdom), the National Research Council (Canada), CONICYT (Chile), the Australian Research Council (Australia), Minist\'{e}rio da Ci\^{e}ncia, Tecnologia e Inova\c{c}\~{a}o (Brazil) and Ministerio de Ciencia, Tecnolog\'{i}a e Innovaci\'{o}n Productiva (Argentina)}.  The position angle of the slit was $25\deg$ east of north so as to align the slit along the major axis of the galaxy.  The effective airmass for the observations ranged from 1.016 to 1.041, making differential refraction effects negligible.  Upon inspection of the pointing frames, it became apparent that the seeing was considerably worse on the second night and the pointing missed the nucleus.  Consequently, we carried out our primary analysis on the three successful observations from the first night.  

The data were reduced and calibrated with the {\it Gemini} package for \texttt{IRAF}, v.~1.11, following standard procedures.  Cosmic rays were cleaned using the \texttt{gscrspec} multi-extension fits wrapper for \texttt{L.A.Cosmic}~\citep{vanDokkum01} and using the default parameters.  Despite the large ($\sim330\arcsec$) length of the slit, some contamination remained from the galaxy even at the ends.  While this was a minor drawback and did not affect the subsequent analysis, it did make full sky line removal impossible.  A set of standard spectroscopic star observations were obtained on the same night using the $5\arcsec$ slit mask, and so we found that slit losses through the $1\arcsec$ slit mask were negligible.

\subsection{VLA Radio Observations}
We observed NGC4178 on 2012 July 14-15 with the NRAO\footnote{The National Radio 
Astronomy Observatory is a facility of the National Science Foundation operated 
under cooperative agreement by Associated Universities, Inc.} Karl G.~Jansky VLA~\citep{perley11} while in its B-configuration (program AS1114, 
12A-093). In the 1.5 hr observing run, a single 18 min scan of the galaxy was 
obtained at C-band while two 18 min scans were obtained at X-band. The target scans 
were bracketed with observations of the point source J1224+0330 for phase 
calibration while 3C~286 was utilized as the primary flux and bandpass calibrator. 
After basic calibration in AIPS, we produced CLEAN images at C-band split into two 
1 GHz wide side-bands centered at 5.0 and 6.0 GHz with off-source rms ($1\sigma$) = 
11.9 $\mu$Jy/beam and 12.6 $\mu$Jy/beam, respectively. The center frequency of the 
X-band image was 9.0 GHz (2 GHz bandwidth) with an off-source rms of 6.7 
$\mu$Jy/beam. The image beamsizes were $1.19\arcsec \times 1.49\arcsec$ (5 GHz), 
$1.03\arcsec \times 1.28\arcsec$ (6 GHz), and $0.66\arcsec \times 0.84\arcsec$ (9 
GHz).

\begin{figure}
\noindent{\includegraphics[width=8.7cm]{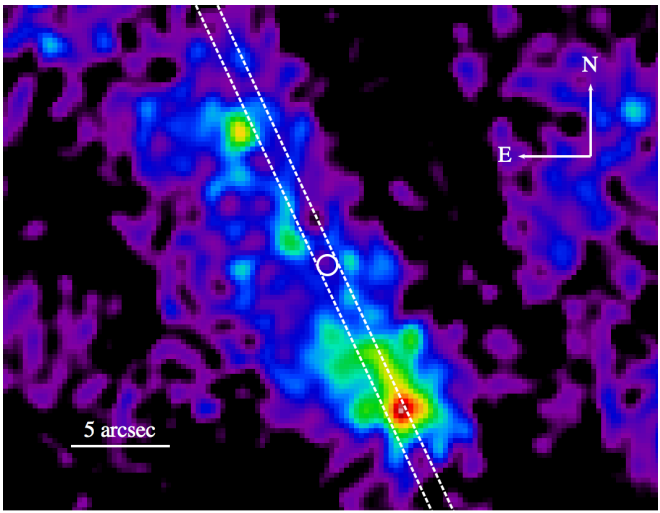}}
\caption{VLA 5 GHz image (beam = $1.49\arcsec\times1.19\arcsec$ at position angle $11.2\deg$) of NGC~4178 showing diffuse emission from the galaxy.  The white circle outlines the {\it Chandra} X-ray source, and the dashed white lines indicate the {\it Gemini} longslit.  $1\arcsec\simeq{160}$ pc.\\}
\label{5GHz_Image}
\end{figure}

In the VLA images, diffuse steep-spectrum emission from the NGC4178 galaxy was 
detected but the Chandra detected nucleus (A) and the three off-nuclear X-ray point 
sources (B, C, and D) reported in \citetalias{secrest12} were not detected. For the nucleus, 
we measured $3 \sigma$ point-source upper limits of $<84.9$ $\mu$Jy (5 GHz), 
$<46.8$ $\mu$Jy (6 GHz), and $<24.9$ $\mu$Jy (9 GHz) at the Chandra X-ray position. 
These values are larger than the off-source rms due to the presence of the NGC4178 
galaxy emission. The corresponding limit from our previous analysis of an archival 
VLA 5 GHz exposure was $<230$ $\mu$Jy \citepalias{secrest12}. For the off-nuclear X-ray 
sources, we measured (5 GHz, 6 GHz, 9 GHz) [$\mu$Jy] upper limits of ($<38.7$, 
$<55.2$, $<25.5$) for B, ($<27.0$, $<36.6$, $<17.1$) for C, and ($<35.7$, $<31.2$, 
$<19.8$) for D. The 5 GHz values are $4.4-5.6 \times$ smaller than determined from 
the archival VLA data~\citepalias[See][]{secrest12}.  Note that the diffuse galactic radio emission is resolved in the $\sim$arcsecond resolution VLA image (Figure~\ref{5GHz_Image}) and exhibits a clumpy appearance due to the CLEAN processing of the data with limited $(u,v)$ coverage.  In this respect, the local peak (66$\mu$Jy/beam) in the new VLA 5 GHz image of the galaxy that we noted in our previous analysis of a sparser archival VLA dataset~\citepalias{secrest12} is noticeably offset (about $1.7\arcsec$) from the {\it Chandra} nucleus and is likely unrelated to the AGN activity.

\section{Results}
\begin{figure}
\noindent{\includegraphics[width=8.7cm]{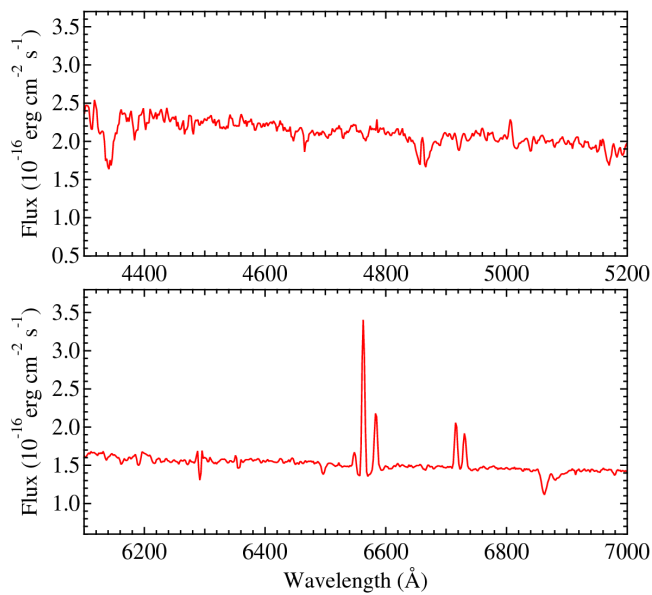}}
\caption{The {\it Gemini} optical spectrum of the central $1\arcsec$ of NGC 4178.\\}
\label{fullSpectrum}
\end{figure}

The calibrated {\it Gemini} spectrum can be seen in Figure~\ref{fullSpectrum}.  Emission line and stellar population fitting were done with \texttt{simplefit}~\citep{tremonti04} using the stellar population model of~\citet{bruzual03}.  One set of 1-dimensional ($1\arcsec$ aperture) and 2-dimensional spectra were produced by performing sky subtraction using the ends of the slit as the sky sample.  The line fluxes for the central $1\arcsec$ are listed in Table~\ref{lineFluxes}.

\begin{center}
\begin{table}
\caption{Line fluxes from the central $1\arcsec$}
\begin{tabular*}{0.48\textwidth}{ c c c } 
Line & Flux$^*$ & EW(\AA)\\[0.5ex] 
\hline \\[0.01ex]
H$\beta$ & $0.33\pm0.02$ & 1.57\\[0.1cm]
[O~III]$\lambda5007$ & $0.18\pm0.02$ & 0.88\\[0.1cm]
H$\alpha$ & $1.31\pm0.01$ & 8.74\\[0.1cm]
[N~II]$\lambda6584$ & $0.41\pm0.01$ & 2.78\\[0.1cm]
[S~II]$\lambda6717$ & $0.36\pm0.01$ & 2.48\\[0.1cm]
[S~II]$\lambda6731$ & $0.26\pm0.01$ & 1.81\\[0.3cm]
\hline 
\end{tabular*}

\begin{flushleft}
$^*$~in units of $10^{-15}$~erg~s$^{-1}$~cm$^{-2}$\vspace{0.05cm}
\end{flushleft}
\label{lineFluxes}
\end{table}
\end{center}

\subsection{The Dynamical Center}
To derive a robust rotation curve, we measure the radial velocity as a function of radius in two ways: 1)  by cross-correlating the spectrum at each spatial position against galaxy templates to determine the velocity, and 2) by fitting directly for the centroids of the H$\alpha$ emission line. For each method, we treat each spatial row of the longslit spectrum independently, but we note that the slit width and seeing are both larger than this pixel size, such that neighboring rows are not independent. For our purposes, it is desirable to assess whether emission line ratios which rise and then fall across a small spatial extent indicate an unresolved AGN source, or arise within a larger resolved region.
 
Under the first method, we follow a similar procedure to that used by SDSS, where the spectrum at each position is transformed into an evenly spaced grid in log wavelength, and then cross-correlated with several similarly-prepared template spectra to determine the redshift (or velocity) of that spectrum with respect to the rest frame. Under the second method, using the cross-correlation result as a first-guess, we fit and subtract off the absorption component of each spectrum (following the procedure described below), and then measure the centroid of a Gaussian fit to the H$\alpha$ line. We note that the two methods yielded consistent curves at all radii for this galaxy.
 
\begin{figure}
\noindent{\includegraphics[width=8.7cm]{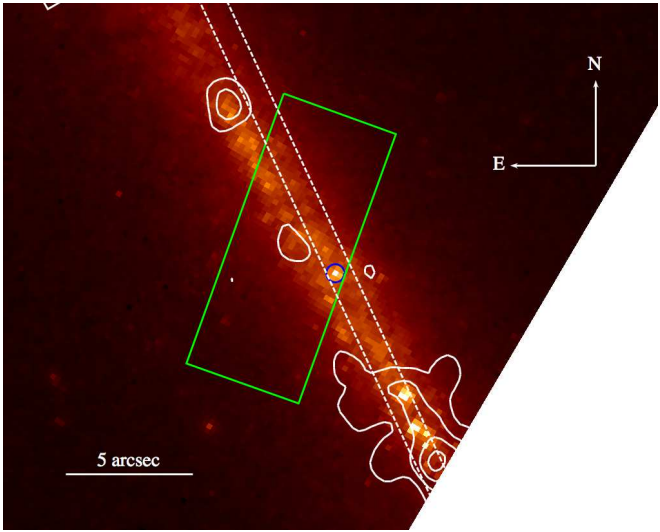}}
\caption{The {\it Hubble Space Telescope} H-band image of the nuclear region of NGC~4178 with the {\it Spitzer} SH slit outlined in green, the {\it Chandra} X-ray source outlined in blue, and the {\it Gemini} longslit outlined in dashed white.  The white contours are the VLA 5 GHz radio contours.  $1\arcsec\simeq{160}$ pc.\\}  
\label{nuclearRegion}
\end{figure}

\begin{figure}
\noindent{\includegraphics[width=8.7cm]{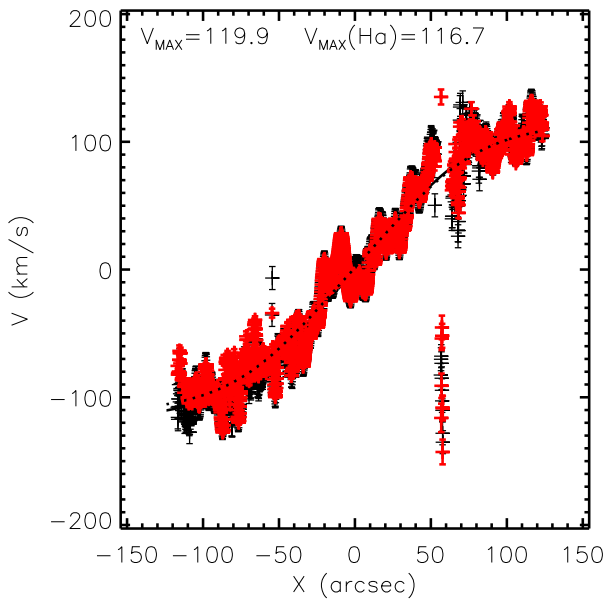}}
\caption{The rotation curve of NGC 4178.  The red points are the H$\alpha$, while the black points are from the absorption line fits.  The x axis is defined relative to the {\it Chandra} nuclear point source.  The uncertainties in the velocities are less than $\sim10\%$, or 10-15~km~s$^{-1}$ for both V$_{\mathrm{MAX}}$ and V$_{\mathrm{MAX}}$(H$\alpha$).  $1\arcsec\simeq{160}$ pc.\\}  
\label{rotation}
\end{figure}

To the raw measured rotation data, we fit a smooth function of the form $V(R)=V_{\rm MAX} (R+\Delta R)/((R+\Delta R)^a+R_s^a)^{1/a}+\Delta V$~\citep{Boehm04,Moran07}, where $R$ is the radius, $a$ and $R_s$ are free parameters that govern the shape of the rotation curve and its turn-over, $\Delta V$ is the offset of the galaxy's central velocity from the given value ($z=0.001248$), and $\Delta R$ is the radius offset . Both $\Delta R$ and $\Delta V$ are left free to vary in the fit, but are constrained to stay within a reasonable range of values (5$\arcsec$ and 75 km s$^{-1}$, respectively). Notably, we find no evidence that $\Delta R$ is significantly different from zero (the uncertainty, as estimated from bootstrapping, was better than $\sim2\arcsec$, or $\sim320$~pc), implying that the presumed center of the galaxy (the location of the {\it Chandra} X-ray source measured in~\citetalias{secrest12}) is also consistent with being its rotational center (Figure~\ref{rotation}).  While the error on the location of the dynamical center appears to preclude us from completely ruling out the possibility that the dynamical center is coincident with the local radio peak $1.7\arcsec$ northeast of the {\it Chandra} source, there are two reasons why this scenario is unlikely.  First, the error in the determination of the dynamical center is along the spatial direction of the slit, which did not cover the local radio peak (See Figure~\ref{nuclearRegion}), so the velocity offset at the location of the radio peak is not known.  Second, the {\it Chandra} source is coincident with the Two Micron All Sky Survey photocenter, which is coincident with a nuclear star cluster~\citepalias[NSC, see Section 3 in][]{secrest12}, making it {\it a priori} more likely that the dynamical center is coincident with the {\it Chandra} X-ray source.

After deriving the best-fit rotation curve, we adjust the spectrum at each spatial position to its rest-frame wavelength scale, and perform all further analysis in the rest frame.

 \subsection{The BPT Diagram}

We employ a modified version of the \texttt{simplefit} code that fits stellar templates and emission lines for each row of the 2-dimensional longslit data.  In order to increase S/N, the spectra were extracted in $1\arcsec$ bins along the spatial direction of the slit, with each consecutive bin shifted by 1 row.  This had the effect of slightly smoothing the data and reducing the noise.  Each extracted spectrum was fit with a linear combination of templates drawn from the~\citet{bruzual03} single stellar population models of varying metallicities, masking out the emission lines.  The best-fitting stellar continuum model was then subtracted from the measured spectrum, creating an emission-line only spectrum where the Balmer emission lines can be measured free of contamination from the underlying stellar continuum.  We then fit Gaussian functions to the emission lines, with the widths and centers of the Gaussian free to vary.

\begin{figure}
\noindent{\includegraphics[width=8.7cm]{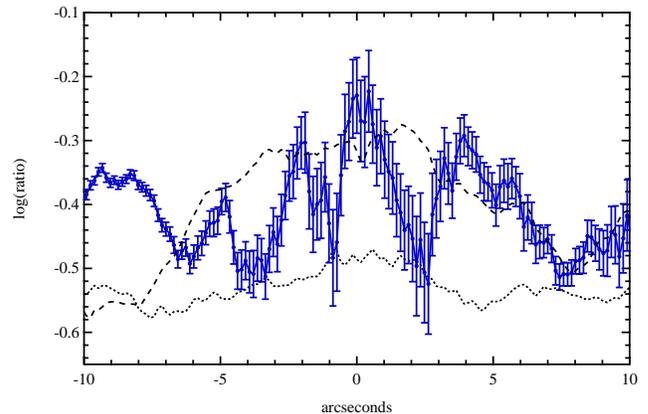}}
\caption{BPT diagnostics as a function of distance from the nuclear source ($1\arcsec\simeq{160}$~pc).  Blue: [O~III]/H$\beta$ with $1\sigma$ error bars.  Dotted: [N~II]/H$\alpha$.  Dashed: [S~II]/H$\alpha$.  The error bars on [N~II]/H$\alpha$ and [S~II]/H$\alpha$ are not significantly larger than the markers, and so have been omitted for clarity.\\}
\label{spatialBPT}
\end{figure}

In addition to  the Balmer lines H$\alpha$ and H$\beta$, we also measure the forbidden lines [O~III]$\lambda5007$, [N~II]$\lambda6548/6584$, and [S~II]$\lambda6717/6731$, which are required t measure metallicity and test for the presence of an AGN. To assess the presence of an AGN, we use the~\citet{kewley01} version of the classic  Baldwin, Phillips \& Terlovich (1981), or BPT, diagnostic, which utilizes the line flux ratios of [O~III]$\lambda5007$ to H$\beta$, [N~II]$\lambda6584$ to H$\alpha$, and [S~II]$\lambda6717/6731$ to H$\alpha$. As a function of radius, we plot these line ratios to assess how rapidly the ionization state of the ISM may change with location in the galaxy~Figure~\ref{spatialBPT}.  While the line ratios vary with radius, at all points their location on the BPT diagram is consistent with originating in HII regions.

\subsection{The Fundamental Plane}
In \citetalias{secrest12}, we used a radio upper limit obtained from archival VLA data combined with the {\it Chandra} X-ray detection to set an upper limit on the nuclear black hole's mass using the so-called fundamental plane of black hole activity \citep{merloni03,falcke04,merloni06,wang06,li08,gultekin09,miller-jones12,bonchi13}.  With the new 5 GHz $3\sigma$ flux density upper limit of $<84.9$ $\mu$Jy and the $0.5-10$~keV luminosity calculated from the scenario in which the AGN is heavily embedded with only a few non-Compton thick openings in which X-ray flux escapes~\citepalias[see Section 3.1 of][]{secrest12}, we find a predicted black hole mass  of $M_{\mathrm{BH}}\lesssim8.4\times10^4$~M$_{\sun}$ using the \citet{miller-jones12} relation ($1\sigma$ scatter = 0.44 dex), which may be more valid than the~\citet{gultekin09} relation for lower-mass systems because the ~\citeauthor{gultekin09} relation was derived for black hole masses greater than $10^6$~M$_{\sun}$ (see Section 4.2 of \citeauthor{miller-jones12}).  This is a considerably lower upper limit than our previous value of $\sim2\times10^{5}$~M$_{\sun}$.  

However, it is important to point out that the~\citeauthor{miller-jones12} relation was derived from the strongly sub-Eddington ($\dot{M}/\dot{M}_{\mathrm{Edd}} < 0.01-0.02$) sample of~\citet{plotkin12}, while significant scatter exists for more highly-accreting objects~\citep[e.g.][]{kording06}, as the AGN in NGC~4178 likely is~\citepalias{secrest12}.  We discuss this issue in the next section.

\section{Discussion}

With the confirmation that the dynamical center of NGC 4178 coincides with the location of the {\it Chandra} X-ray point source, an important property of the AGN in this galaxy has been confirmed.  While bulge-hosting galaxies usually have dynamical centers very clearly aligned with their photocenters, bulgeless galaxies, especially bulgeless galaxies with dusty bars such as NGC 4178, have much less obvious centers, and several sources within the bar of NGC 4178 could be considered candidates for the dynamical center, such as one of the two IR-bright lobes detected by {\it Spitzer}~\citepalias[see Figure 1 of][]{secrest12}.  This analysis highlights the value of using longslit spectroscopy on this unique class of objects, despite the fact that optical emission lines originating from the narrow line region (NLR) may either be too absorbed or too weak to use as an AGN diagnostic by themselves.

It is not clear from the data alone whether it is heavy absorption that prevents NGC 4178 from being optically classified as an AGN or if it is simply that the AGN is too weak to be detected above surrounding star formation.  By comparing the X-ray properties of the AGN with the bolometric luminosity predicted from the [Ne~V] 14.3~$\micron$ emission, \citetalias{secrest12} found that the AGN in NGC 4178 is indeed likely to be heavily obscured, and so we should therefore not be surprised that this galaxy shows no signs of an AGN at optical wavelengths.  Even with heavy obscuration, we may be able to detect an increase in ionization parameter near the AGN, giving some clues as to degree of absorption involved.  One advantage of longslit spectroscopy is that it has allowed us to examine the ionization state of the ISM as a function of radius from the AGN.  Of the three ionization parameters used in~\citet{kewley01}, all three appear to rise near the galactic center, but only [O~III]/H$\beta$ line flux ratio appears to increase sharply enough at the galactic center to possibly be considered local to the AGN (Figure~\ref{spatialBPT}).  However, the magnitude of the rise in the line flux ratio at the center is comparable to that of the nearby off-nuclear variations. The rise in the line flux ratio at the center cannot alone be considered unambiguous evidence for the AGN.

\begin{figure}
\noindent{\includegraphics[width=8.7cm]{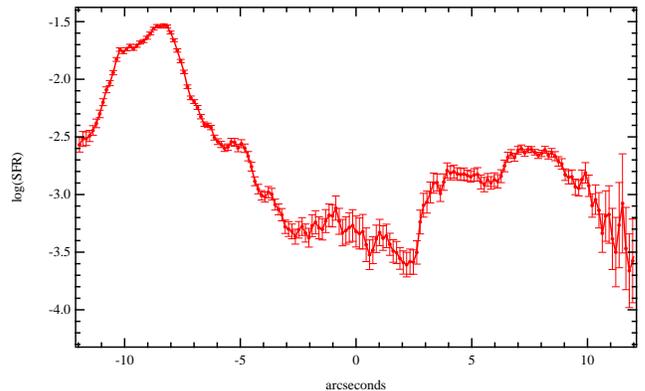}}
\caption{SFR (M$_{\sun}$~yr$^{-1}$) as a function of angular distance from the nuclear source.  Error bars are 1$\sigma$. $1\arcsec\simeq{160}$ pc.\\}
\label{spatialSFR}
\end{figure}

There is, however, some amount of scatter in the $L_{\mathrm{[Ne V]}}/L_{\mathrm{bol}}$ relation (0.53 dex), and a portion of this scatter is likely due to [Ne~V] contamination by heavy starburst activity~\citep[e.g.][]{veilleux05}.  In order to explore this possibility, we plot the star formation rate (SFR) as a function of distance from the nuclear source in Figure~\ref{spatialSFR} using the~\citet{kennicutt98} relationship between the H$\alpha$ luminosity and the SFR and using the Charlot~\&~Fall law~\citep{charlotfall00} to empirically correct for extinction assuming optically-thick (Case B) recombination.  The IR-bright lobes detected by {\it Spitzer} at $\sim7\arcsec$ away from the AGN are clearly visible as large increases in the SFR.  However, within $\sim3\arcsec$ of the nucleus, the SFR hovers at around $10^{-3}$~M$_{\sun}$~yr$^{-1}$, indicating that there is no significant starburst activity near the nucleus, as is implied by the absence of significant Pa-$\alpha$ emission noted by \citetalias[][Figure 2 therein]{secrest12}.  Being an extremely late-type, and therefore very dusty galaxy, it is natural to ask if there might be a deeply buried starburst contributing to the [Ne~V] emission that could be obscured even in the near-IR.  To explore this possibility, we used the tight relationship between $L_{\mathrm{[Ne~II]+[Ne~III]}}$ and SFR~\citep{hoKeto07}, which is largely insensitive to differences in metallicity, to calculate the SFR at the nucleus of NGC~4178 under the possible scenario in which the [Ne~II] + [Ne~III] emission is due to a buried starburst and not to the presence of an AGN.  Using the best-fit ionization fraction values of \citeauthor{hoKeto07}, we derive a SFR of $\sim4\times10^{-2}$~M$_{\sun}$~yr$^{-1}$.  As this number is consistent with the rate predicted by the H$\alpha$ emission within the margin of error, we conclude that there is not a buried starburst contributing to the [Ne~V] emission in NGC~4178.

The new upper limit to the 5 GHz radio luminosity further constrains the upper limit on the black hole mass to within the same order of magnitude as the Eddington lower mass limit implied by the [Ne V] luminosity.  This would seem to corroborate the scenario of an IMBH deeply embedded in a heavy absorbing material and accreting at a high rate.  However, it is important to note that the {\it Chandra} point source in NGC~4178 is thus inferred to be in the `high-soft' state of black hole X-ray emission~\citepalias[see section 3.1 of][]{secrest12}, and so the radio and X-ray luminosities may be only weakly dependent on the mass of the nuclear black hole~\citep[for details, see][]{wang06,li08}.  The upper limit on the mass implied by the deeper VLA measurements should therefore be considered tentative~\citep[Although see][for an example of an AGN accreting very close to the Eddington limit with a mass accurately predicted by the fundamental plane]{gliozzi10}. 

\section{Summary and Conclusions}

We have analyzed {\it Gemini} longslit optical spectroscopy of NGC~4178, as well as deeper VLA radio observations.  By performing a full spatial spectroscopic characterization of the galaxy, as well as more tightly constraining its radio properties, we have found the following results:

\begin{enumerate}
\item{The dynamical center of NGC~4178 is coincident with the location of the {\it Chandra} X-ray point source discovered by~\citetalias{secrest12}, providing further support for the presence of an AGN in this galaxy, as well as confirming that this galaxy contains a nuclear star cluster.}
\item{While the ionization state of the ISM near the nucleus appears to increase, NGC~4178 remains classified as optically normal, further underscoring the need for the penetrative power of mid-IR and X-ray observations in finding buried or energetically weak AGNs in this class of galaxy.}
\item{From new, deeper constraints on the 5 GHz radio luminosity, the upper limit on the nuclear black hole's mass appears to be near the Eddington limit implied by its mid-IR properties, lending support to the notion that this object is deeply embedded and accreting at a high rate.  The uncertainty in applying the fundamental plane to highly-accreting black holes, however, means that we should consider this to be a tentative result.}

\end{enumerate}

\section{Acknowledgements}

We thank the referee for their very thorough review and insightful comments that has significantly improved this paper.  N.\init J.\init S.~and S.\init S.~gratefully acknowledge support by the {\it Chandra} Guest Investigator Program under NASA Grant G01-12126X.  C.\init C.\init C.~was supported at NRL by a Karles' Fellowship and NASA DPR S-15633-Y.  M.\init G.~acknowledges financial contribution from the agreement ASI-INAF I/009/10/0.\\

\end{document}